\documentclass[epj]{svjour}
\usepackage{graphicx}
\begin{document}

\title{Influence of optical aberrations in an atomic gyroscope}
\author{J\'{e}r\^{o}me Fils\inst{1}, Florence Leduc\inst{1}, Philippe Bouyer\inst{2},
David Holleville\inst{1}, No\"{e}l Dimarcq\inst{1}, Andr\'{e}
Clairon\inst{1} \and Arnaud Landragin\inst{1}
}                     

\institute{BNM-SYRTE, UMR 8630, Observatoire de Paris, 61 avenue
de l'Observatoire, 75014 Paris, France \and Laboratoire Charles
Fabry, UMR 8501, Centre Scientifique d'Orsay, $b\hat{a}t.$ 503, BP
147, 91403 Orsay, France}
\date{Received: date / Revised version: date}
%

 \abstract{ In atom interferometry based on light-induced diffraction, the optical aberrations of the
  laser beam splitters are a dominant source of noise and systematic effect.
 In an atomic gyroscope, this effect is dramatically reduced by the use of two atomic sources. But it
 remains critical while coupled to fluctuations of atomic trajectories, and appears as a main source of
  noise to the long term stability. Therefore we measure these contributions in our setup, using cold Cesium atoms and stimulated Raman transitions.
} 

\maketitle
\section{Introduction}
\label{intro} Since the pioneering demonstrations of
interferometry with de Broglie atomic waves using resonant light
\cite{BordePTB,Bragg} and nanofabricated structures \cite{Prit91}
as atomic beam splitters, a number of new applications have been
explored, including measurements of atomic and
molecular properties, fundamental tests of quantum mechanics, and
studies of various inertial effects \cite{Berman}. Using atom
interferometers as inertial sensors is also of interest for geophysics,
tests of general relativity \cite{Hyper}, and inertial guidance
systems.

Atom interferometers based on light-induced beam splitters have
already demonstrated considerable sensitivity to inertial forces.
Sequences of optical pulses generate the atom optical elements
(e.g., mirrors and beam splitters) for the coherent manipulation
of the atomic wave packets \cite{Borde91}. The sensitivity and
accuracy of light-pulse atom interferometer gyroscopes
\cite{Gustavson00}, gravimeters \cite{Peters99} and gravity
gradiometers \cite{Snadden98} compare favorably with the
performances of state-of-the-art instruments. Furthermore, this
type of interferometer is likely to lead to a more precise direct
determination of the fundamental constant $\alpha$ from the
measurement of $\hbar/M$ \cite{wicht2001}. In the case of rotation
measurements, the sensitivity reaches that of the best laboratory
ring laser gyroscope \cite{Stedman}. Indeed the Sagnac phase
shift, proportional to the total energy of the interfering
particle, is much larger for atoms than for photons. This
compensates for the smaller interferometer area and the lower
flux.

 In this paper we focus on the effect of the fluctuations of
the atomic trajectory, which might affect the long term stability
of atomic gyroscopes when coupled with local phase variations
induced by optical aberrations. We will introduce this problem in
paragraph 2 and illustrate it quantitatively in the case of our
setup in paragraph 3.

 Our experiment consists in an almost complete inertial
measurement unit \cite{Yver03}, using cold Cesium atoms that
enable for a drastic reduction of the apparatus dimensions while
reaching a sensitivity of $30$ nrad.s$^{-1}.$Hz$^{-1/2}$ to
rotation and 4x10$^{-8}$ m.s$^{-2}.$Hz$^{-1/2}$ to acceleration.
Its operation is based on recently developed atom interference and
laser manipulation techniques. Two interferometers with
counter-propagating atomic beams discriminate between rotation and
acceleration \cite{Gustavson98}. Thanks to the use of a single
pair of counter-propagating Raman laser beams, our design is
intrinsically immune to uncorrelated vibrations between the three
beam splitters, usually limiting such devices. This configuration
is made possible by the use of a reduced launch velocity, inducing
a reasonable interaction time between the pulses. However, as any
atomic gyroscope, our sensor's scheme remains sensitive
 to local phase variations, a limitation that
has already been encountered in optical atomic clocks \cite{Trebst01}.

\begin{figure}
  \centering
  \includegraphics[width=3in]{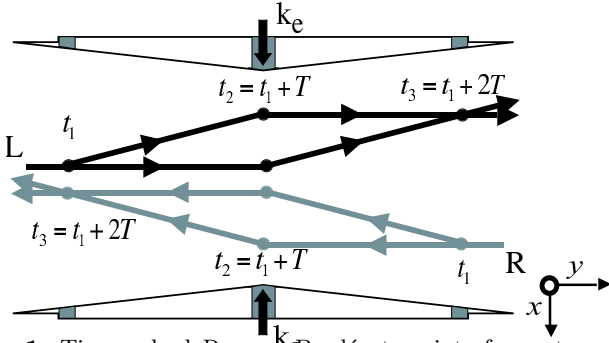}
\caption{Time-pulsed Ramsey-Bord\'{e} atom interferometer using
stimulated Raman transitions induced
  by two counter-propagating laser beams of wave vectors ${\bf k}_{e}$ and ${\bf k}_{g}$. Cesium atoms
   are launched on the same trajectory but in opposite directions with velocities ${\bf v}^{L,R}=\{0,\pm v_{y},v_{z}\}$,
    from right to left (R) and left to right (L).  The interactions with light pulses occur at times $t_{i=1,2,3}$
     at three different locations. The detection consists in measuring the probability of presence in each output port after the last pulse.}
\label{fig:1}
\end{figure}

\section{Principle}
\label{prin}

 We first briefly review the basic light-pulse
method in the case of a symmetric Ramsey-Bord\'{e} interferometer
scheme \cite{Borde02}, where three travelling-wave pulses of light
resonantly couple two long-lived electronic states.The two-photon
stimulated Raman transitions between ground state hyperfine levels
are driven by two lasers with opposite propagation vectors  ${\bf
k}_{e}$ and ${\bf k}_{g}$ (${\bf k}_{e}\simeq{\bf -k}_{g}$).
First, at $t=t_1$ a beam splitting pulse puts the atom into a
coherent superposition of its two internal states. Because of
conservation of momentum during the atom-light interaction, this
pulse introduces a relative momentum $\hbar{\bf k}_{\rm
eff}=\hbar{\bf k}_{g}- \hbar{\bf k}_{e}$ between the atomic wave
packets corresponding to each state. These wave packets drift
apart for a time $T$, after which a mirror pulse is applied at
$t_2=t_{1}+T$ to redirect the two wave packets. After another
interval of duration $T$, the wave packets physically overlap, and
a final beam splitting pulse recombines them at $t_3=t_{1}+2T$.
The measurement of the probabilities of presence in both internal
states at the interferometer output leads to the determination of
the difference of accumulated phases along the two paths. In
general, atoms are launched with a velocity $\bf v$ so that each
stimulated Raman transition occurs at a particular position
$\left\{x_i,y_i,z_i\right\}_{i=1,2,3}$ that can be evaluated from
the classical trajectories associated with the atomic wave packets
\cite{Antoine03}, as shown fig. \ref{fig:1}. In our setup, Raman
laser beams propagate in the \emph{(Ox)} direction and atoms are
launched in the $(y,z)$ plane. We define ${\bf
u}_{i}$=$\left\{y_i,z_i\right\}$ the atomic cloud positions in
this plane at time $t_i$.

In the absence of any external forces, atoms initially prepared in
a particular state ($^6S_{1/2},F=3,m_F=0$ in the present setup)
will return to this state with unit probability. A uniform
external acceleration or rotation induces a relative phase shift
between the interfering paths. This phase shift modifies the
transition probability between the two Cesium internal states
$^6S_{1/2},F=3,m_F=0$ and $^6S_{1/2},F=4,m_F=0$ (noted $|3\rangle$
and $|4\rangle$ in the following). Hence the transition
probability measurement leads to the determination of the phase
shift and finally the evaluation of the perturbing forces.

It can be shown that the only contribution to the phase shift
results from the interaction with the laser light fields
\cite{Antoine03}. In the limit of short, intense pulses, the
atomic phase shift associated with a transition
$|3\rangle\rightarrow |4\rangle$ (resp. $|4\rangle\rightarrow
|3\rangle$) is +$\phi_i$ (resp. -$\phi_i$), where $\phi_i$ is the
phase difference between the two Raman laser beams. We then find
that the transition probability from $|3\rangle$ to $|4\rangle$ at
the exit of the interferometer is simply
$\frac{1}{2}\left[1-\cos(\Delta\phi) \right]$ where $\Delta\phi =
\phi_1- 2\phi_2+\phi_3$. The three quantities correspond to the
phase imparted to the atoms by the initial beam splitting pulse,
the mirror pulse, and the recombining pulse where
$\phi_i=\phi_{g}\left(u_i,t_i\right)-\phi_{e}\left(u_i,t_i\right)={k}_{\rm
eff}\cdot{x}_{i}+\Phi({\bf u}_{i})$. The sensitivity to rotation
and acceleration arises from the first term ${k}_{\rm
eff}\cdot{x}_{i}$ and simplifies to $\Delta\phi_{\rm acc}=
a_{x}k_{\rm eff}T^2$ and $\Delta\phi_{\rm rot} = -2 k_{\rm eff}
v_{y}\Omega_{z} T^2$ for the present setup. The phase $\Phi({\bf
u}_{i})$ for the pulse at time $t_{i}$ corresponds to the local
phase in the $(y,z)$ plane due to wavefront distortions of both
laser beams\footnote{The interferometer is also sensitive to time
fluctuations of the Raman laser phases \cite{Yver03}. These
fluctuations are identical for the two interferometers and
disappear from the rotation signal. They will be neglected in this
paper.}. It induces a residual phase error at the exit of the
interferometer $\delta\Phi = \Phi({\bf u}_{1})-2\Phi({\bf
u}_{2})+\Phi({\bf u}_{3})$.

Acceleration cannot be discriminated from rotation in a single
atomic beam sensor, as stated above. This limitation can be
circumvented by installing a second, counter-propagating, cold
atomic beam (fig. \ref{fig:1}) \cite{Gustavson98}. When both
atomic beams perfectly overlap, the area vectors
for the resulting interferometer loops have opposite directions.
The corresponding rotational phase shifts $\Delta\phi_{\rm
rot}$ have opposite signs while the acceleration phase shifts
$\Delta\phi_{\rm acc}$ are identical. Consequently, acceleration
is calculated by summing the two interferometer's phase shifts:
$\Delta\phi_{+}\sim 2\Delta\phi_{\rm acc}$; while taking the
difference rejects the
contribution of uniform accelerations so that $\Delta\phi_{-}\sim
2\Delta\phi_{\rm rot}$. In addition, the residual phase error
$\delta\Phi$ vanishes in $\Delta\phi_{-}$, but remains in $\Delta\phi_{+}$
as an absolute phase bias $2\times\delta\Phi$.

However, an imperfect overlapping of the two counter-propagating
wavepackets trajectories might lead to an imperfect common mode
rejection of the residual phase error in $\Delta\phi_{-}$. Thus, a
phase bias $\delta\Phi_{-}=\delta\Phi^{L}-\delta\Phi^{R}$ will
appear, where the notations $^{L}$ and $^{R}$ concern the left and
right atom interferometers. While the phase bias
$\delta\Phi_{+}\simeq2\times\delta\Phi$ depends on the local value
of the phase at the average position ${\bf r}_{i}= \frac{{\bf
u}_{i}^{L}+{\bf u}_{i}^{R}}{2}$, the phase bias $\delta\Phi_{-}$
depends on  the local phase gradient at the average position ${\bf
r}_{i}$ with the position offset $\delta{\bf r}_{i}={\bf
u}_{i}^{L}-{\bf u}_{i}^{R}$:
\begin{equation}\label{DPAB2b}
  \begin{array}{ccc}
\delta\Phi_{-}  & = & \boldmath{\nabla}\Phi({\bf r}_{1})\cdot
\delta{\bf
r}_{1} - 2\boldmath{\nabla}\Phi({\bf r}_{2})\cdot \delta{\bf r}_{2}\\
& & +
\boldmath{\nabla}\Phi({\bf r}_{3})\cdot \delta{\bf r}_{3}.\\
  \end{array}
\end{equation}
Equation \ref{DPAB2b} shows that uncorrelated fluctuations of the
wavepackets trajectories from shot to shot causes fluctuations of
the phase bias, which amplitude depends on the local wavefront
slope of the phase. If we consider a perfect control of the launch
velocity\footnote{We can reach a stability of $10^{-4} $m.s$^{-1}$
or better from shot to shot thanks to the moving molasses
technique \cite{molasses}.}, fluctuations of trajectories are only
due to fluctuations of the initial positions of the atomic clouds.
Consequently, we can consider $\delta {\bf r}_1=\delta {\bf
r}_2=\delta {\bf r}_3$. The phase fluctuation is then simply
proportional to the product of the fluctuations of the cloud
initial position $ {(y_0, z_0)}$ with the phase gradients
$\Delta\Phi_i$. As the phase gradients are time-independent, the
Allan variance of the phase $\sigma^2_{\delta\Phi_{-}}$  is
simply:
\begin{equation}\label{Allanposition}
 \begin{array}{ccc}
  \sigma^2_{\delta\Phi_{-}}=\sigma^2_{y_0}.\left[\partial_y\left(\Phi\left({\bf
r}_{1}\right)-2\Phi\left({\bf r}_{2}\right)+\Phi\left({\bf
r}_{3}\right)\right)\right]^2 \\
+
 \sigma^2_{z_0}.\left[\partial_z\left(\Phi\left({\bf
r}_{1}\right)-2\Phi\left({\bf r}_{2}\right)+\Phi\left({\bf
r}_{3}\right)\right)\right]^2\
 \end{array}
\end{equation}
where $\sigma^2_{y_0}$ and $\sigma^2_{z_0}$ are the Allan variances of
the initial horizontal and vertical positions. Eq. \ref{Allanposition} shows
 that the fluctuations of the clouds initial positions, as well as the wavefront
 quality of the Raman beams, have to be systematically investigated in atomic gyroscopes in order to estimate how it affects its performances.
\section{Experimental results}
\label{exp}
\begin{figure}
  \centering
  \includegraphics[width=3in]{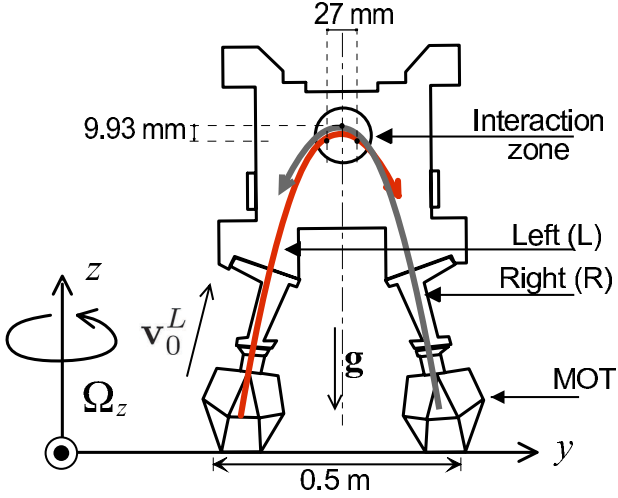}
\caption{Front view of our gyroscope; the interaction zone is
located near the top of
  the atomic trajectories. Atoms
  are launched symmetrically at initial velocity  $v_0 = 2.4 $ m.s$^{-1}$, making an angle
  of $82^o$ with the horizontal axis. The enclosed
  oriented areas are equivalent to their projections on the (Oxy) plane.}
\label{fig:2}
\end{figure}

 In our setup, the atomic sources
are clouds of Cesium atoms, cooled in magneto-optical traps and
launched with a parabolic flight (fig. \ref{fig:2}). As the initial angle reaches $82^o$, and the
launch velocity $2.4 \ $m.s$^{-1}$, the horizontal velocity $v_y$
is $0.3\ $m.s$^{-1}$.
The single pair of Raman laser beams propagates along
the x-axis and is switched on three times at the top of the atomic
trajectories. If the three pulses are symmetric with respect to the trajectory apogees, the interferometer oriented
enclosed areas are equivalent to their flat horizontal
projections: the oriented vertical projection is naught.  The time delay between pulses is typically
$45$ ms. The positions of the atoms during the three Raman pulses are given in fig. \ref{fig:2}.

\begin{figure}
 \centering

  \includegraphics[width=3in]{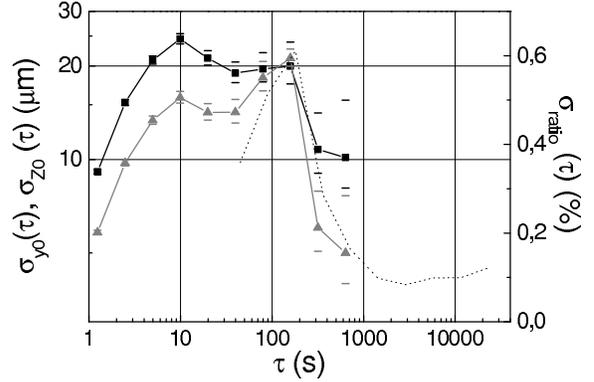}
\caption{Allan standard deviations of the horizontal (black
squares) and vertical (grey triangles) MOT positions as a
  function of the integration time $\tau$, plotted in log-log scale. On the right axis the Allan standard deviation of the intensity ratio of MOT
  cooling lasers is plotted in dashed line as a function of the integration time $\tau$.}
\label{fig:3}       
\end{figure}

In order to investigate the fluctuations of the atomic initial positions from shot to shot,
we image one of the two clouds. The cycling sequence takes about 1.3 s and consists on a trap
phase of 500 ms, a molasses phase of 20 ms, a launching phase of 2 ms and a waiting time phase
 of 800 ms needed to process the image: download of the image, subtraction of a background image
  and determination of the cloud barycenter position in y- and z-axes. The image is taken just
   after turning off the trap magnetic field, at the end of the molasses phase. We calculate the
    Allan standard deviations \cite{Allan} of the
barycenter horizontal and vertical positions (fig. \ref{fig:3}) from a one hour acquisition.
Two peaks, appearing after 10 s and 150 s of integration time, are
characteristic of fluctuations of periods equal to 20 s and 300 s.
After about 10 min integration (630 s), the position standard
deviations reach 10 $\mu$m and 5 $\mu$m in the horizontal and vertical directions
respectively. This dissymmetry is consistent with the magnetic field
gradient configuration, which is twice higher on the Z-direction.
The long-term variations are due to fluctuations of the MOT
cooling lasers intensity ratio, which Allan standard deviation is
plotted in fig. \ref{fig:3}. We see again the oscillation of
period 300 s, appearing for 150 s integration time. We analyze
this as the period of the air conditioning, creating temperature
variations on the fibre splitters delivering the cooling lasers.

This result has to be coupled to the optical aberrations of the
Raman lasers. The main contribution to these aberrations comes
from the vacuum windows used for the Raman laser beams, which
clear diameter is 46 mm. They have been measured with a Zygo
wavefront analyzer, which gives the laser phase distortion created
by the windows. This distortion is projected on the Zernike
polynomial base \cite{Zer}. As our atomic clouds are about 2 mm
wide, the decomposition is pertinent only up to the 36th
polynomial. Indeed, the upper numbers correspond to high spatial
frequencies, so that their effect will be smoothed by averaging on the atomic cloud dimensions.
 To reduce the stress on the vacuum windows, essentially due to the mounting, they were glued in place.
  Thanks to this method, the wavefront quality
reaches $\lambda/50 $ rms over the whole clear diameter of 42 mm.

\begin{figure}
 \centering
  \includegraphics[width=3in]{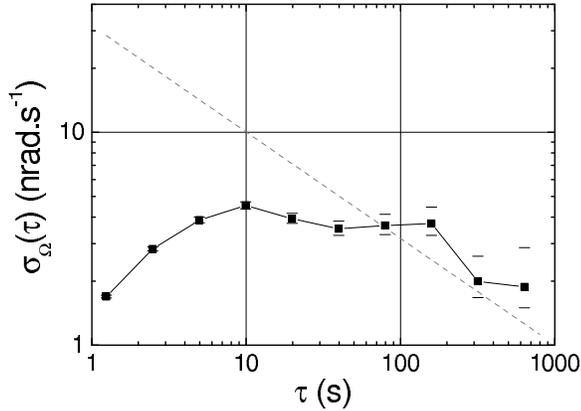}
\caption{Allan standard deviation of the rotation measurement,
taking into account the optical aberrations when coupled with position
  fluctuations. The dashed curve shows the quantum projection
  noise limit, indicating that the optical aberrations may affect the gyroscope performances at long term.}
\label{fig:4}
\end{figure}

The wavefront measurement allows for evaluation of the
atomic phase shift fluctuations due to the coupling between
aberrations and position fluctuations using eq.\ref{Allanposition} assuming that the
two sources are uncorrelated. Their relative position fluctuations
are $\sqrt{2}$ times greater than these observed for one source.
The contribution of this phase fluctuations to the Allan
standard deviation of the rotation rate measurement is shown in fig.
\ref{fig:4}. We compare it with the ultimate stability of our
gyroscope, given by the quantum projection noise.
It is estimated to $30/\sqrt{\tau}\ $nrad.s$^{-1}$
($\tau$ is the integration time) from the ultimate signal-to-noise ratio obtainable with $10^{6}$
atoms.

 The rotation noise induced by position fluctuations has a
significant contribution for integration times larger than 100 s.
 At the present stage of the experiment, this limitation is due
 to the high temperature sensitivity of the fibre splitters.
 This could be the main limitation of
the gyroscope performances.
\section{Conclusion}
\label{concl} In the present paper we studied the stability of a
cold atom gyroscope based on two symmetrical Ramsey-Bord\'{e}
interferometers, with respect to optical phase inhomogeneity.
Instability due to aberrations is not a
specific problem induced by Raman transitions, but concerns every
type of atom interferometer using light beam splitters.
 We showed that the coupling between wavefront distortions of these lasers and fluctuations of the
atomic trajectory becomes predominant at long term, despite a
wavefront quality of $\lambda/50 $ rms obtained thanks to glued
windows. In our setup, atomic trajectory fluctuations are mainly
due to fluctuations of the intensity ratio of the MOT cooling
lasers, induced by the fibre splitters used for their generation.
\newline
However several improvements may render their contribution
negligible:
\newline - reduce the atomic trajectory fluctuations, by using discrete optical couplers for the MOT instead of
the present fibre splitters,
\newline - minimize the number of optics which contribute to the interferometer instability. This can be done by including the Raman laser beam
imposition optics in the vacuum chamber, in order to remove the
aberrations due to the vacuum windows, or by minimizing the number
of non-common optics for the two Raman lasers, since only the
phase difference between the lasers is imprinted on the atomic
phase shift.

Such techniques open large improvement possibilities, which will
be confirmed directly on the long-term stability measurement of
the atomic signal in our interferometer setup.

\section{Acknowledgements}
\label{acknow} The authors would like to thank DGA, SAGEM and CNES
for supporting this work, Pierre Petit for the early stage of the
experiment and Christian Bord\'{e} for helpful discussions. They
also thank Thierry Avignon and Lionel Jacubowiez from SupOptique
for their help in the wavefront measurement.

\end{document}